\documentclass[aip,graphicx]{revtex4-1}
\usepackage[cp1251]{inputenc}
\usepackage[T2A]{fontenc}
\usepackage[english]{babel}
\usepackage{amssymb,latexsym,amsmath,amscd}
\usepackage{graphicx,color,framed}
\begin{document}
\selectlanguage{english}

\title{Statistical description of co-nonsolvency suppression at high pressures}

\author{\firstname{Yu.~A.} \surname{Budkov}}
\email[]{ybudkov@hse.ru}
\affiliation{Tikhonov Moscow Institute of Electronics and Mathematics, School of Applied Mathematics, National Research University Higher School of Economics, Tallinskaya st. 34, 123458 Moscow, Russia}
\author{\firstname{ A.~L.} \surname{Kolesnikov}}
\email[]{kolesnikov@inc.uni-leipzig.de}
\affiliation{Institut f\"{u}r Nichtklassische Chemie e.V., Universit\"{a}t Leipzig, Permoserstr. 15, 04318 Leipzig, Germany}

\begin{abstract}
We present an application of Flory-type self-consistent field theory of the flexible polymer chain dissolved in the binary mixture of solvents to theoretical description of co-nonsolvency. We show that our theoretical predictions are in good quantitative agreement with the recently published MD simulation results for the conformational behavior of a Lennard-Jones flexible chain in a binary mixture of the Lennard-Jones fluids. We show that our theory is able to describe co-nonsolvency suppression through pressure enhancement to extremely high values recently discovered in experiment and reproduced by full atomistic MD simulations. Analysing a co-solvent concentration in internal polymer volume at different pressure values, we speculate that this phenomenon is caused by the suppression of the co-solvent preferential solvation of the polymer backbone at rather high pressure imposed. We show that when the co-solvent-induced coil-globule transition takes place, the entropy and the enthalpy contributions to the solvation free energy abruptly decrease, while the solvation free energy remains continuous.
\end{abstract}

\maketitle
\section{Introduction}
Co-nonsolvency is a phenomenon of polymer insolubility in a mixture of two good solvents. For dilute polymer solutions it can be described as a re-entrant coil-globule-coil transition of a single polymer chain, where the conformational transitions are triggered by varying the co-solvent concentration. Thus, polymer solutions exhibiting co-nonsolvency, are a good example of smart (or environmentally driven) systems. The latter have attracted great attention of researchers due to their non trivial nature and great potential in industrial applications. A typical system for the investigation of co-nonsolvency is PNIPAM in a water/methanol mixture \cite{Wolf78,Shild,Winnik,Zhang,Walter2012,Maekawa,Bischofberger,Hofmann1,Osaka2012,Hofmann2,Wang2012,Bischofberger2014,Mukherji_2016, Schroer_2016,Kyriakos_2016,Vansco_2017,Tanaka2008,Walter2012,Mukherji2013,Kremer2014,Mukherji2015, Kremer2015,Van-der-Vegt2015_2,Van-der-Vegt2017,Freed2015,Freed2016,Sapir2015,Sapir2016,Pica_2016,
Mukherji_2017,Van-der-Vegt_2017,Budkov_conons}. Many studies confirm that the effects of mixed solvents play an important role in different areas of chemical technology.
As is well known, the PNIPAM aqueous solution has a low critical solution temperature (LCST), so the temperature increase always leads to the polymer chain collapse \cite{Bekiranov_1993,Simmons_2008}. However, alcohol additives to the solution may
shift the LCST significantly \cite{Bischofberger2014,Hofmann2}. Nevertheless, it is important to mention that the LCST itself is not the matter of discussion of works \cite{Bischofberger2014,Hofmann2}, but rather the reduction of the LCST.

In recent literature, there are two main points of view on the co-nonsolvency thermodynamic nature. By using the Molecular dynamic (MD) simulations, the authors of works \cite{Mukherji2013,Kremer2014,Mukherji2015} show that a flexible polymer Lennard-Jones (LJ) chain undergoes a reversible coil-globule-coil transition in the mixture of two LJ fluids, when the concentration of one of the solvents increases. The authors performed both all atom and suitably parameterized coarse grained simulations. Both solvents were chosen as good ones with respect to the polymer chain. The MD simulations demonstrate a quantitative agreement with the experimental data of PNIPAM in aqueous methanol \cite{Zhang}.  Moreover, the authors suggest a simple lattice model based on the assumption that co-solvent molecules can be adsorbed on the polymer backbone creating "bridges" between two non-neighboring monomers. Therefore, the authors interpret co-nonsolvency as a pure enthalpic effect.
The authors also demonstrated that by adding alcohol to water the solvent mixture becomes an effectively better solvent even though the polymer chain collapses.

Another theoretical explanation of co-nonsolvency was suggested in paper \cite{Van-der-Vegt2015_2}. In ref. \cite{Van-der-Vegt2015_2} the authors used the fully atomistic MD simulation to investigate the physical mechanism of co-nonsolvency. The authors do not discuss the reduction of the LCST, but rather the transition itself. Because of that they give more weight to the LCST characteristic entropy contribution. They show that co-nonsolvency is driven by the balance between the enthalpy and entropy contributions to the solvation free energy. They stressed out especially that this effect is chemistry-specific, and thus is not a generic phenomenon. As a result, it is shown that during preferential binding of methanol with PNIPAM the energetics of electrostatic, hydrogen bonding, or bridging-type interactions with the globule has been found to play no role. Recently, it has been shown by the same authors that polymer hydration is the determining factor for PNIPAM collapse in the co-nonsolvency regime. In particular, it is shown that methanol frustrates the ability of water to form hydrogen bonds with the amide proton and therefore causes polymer collapse \cite{Van-der-Vegt2017}. Thus, ref. \cite{Van-der-Vegt2017} actually derives a microscopic argument for the preferred coordination of alcohol with PNIPAM as claimed by authors of Refs. \cite{Mukherji2013,Kremer2014,Mukherji2015}.

The authors of the recent research \cite{Hofmann1} have investigated the pressure influence on co-nonsolvency in the aqueous methanol solution of PNIPAM. They have shown that a pressure increase to extremely high values suppresses co-nonsolvency, resulting in the independence of the PNIPAM coil conformation from the solvent composition. The authors have qualitatively interpreted their experimental results, assuming that the pressure increase compensates polymer hydrophobicity, leading to polymer chain expansion \cite{Budkov2}. In the work of de Oliveira et al. \cite{Kremer2015}, the same effect has been reproduced by the full atomistic MD simulation of PNIPAM in aqueous methanol. The simulations provided detailed information through Kirkwood-Buff integrals. The authors relate the high pressure co-nonsolvency suppression to the disappearing of the preferential binding of the co-solvent versus a solvent with a polymer backbone. It is worth noting that the pressure effect on LCST of the polyethylene oxide aqueous solution was theoretically analyzed within a Flory-type model in ref. \cite{Bekiranov_1993}. The authors of ref. \cite{Bekiranov_1993} modify the standard free energy of the Flory theory, taking into account the hydrogen bonding of monomers with water molecules. Thereby, as it could be seen, the pressure effect on the LCST is related to the hydrogen bonding in polymer solutions and being not generic.

Recently, the authors of this communication have formulated the Flory-type self-consistent field theory based on the modern liquid-state theory of the flexible polymer chain in the mixed solvent and applied it to the co-nonsolvency description \cite{Budkov_conons}. It was confirmed that co-nonsolvency could be obtained within the theory taking into account only the universal Van der Waals and excluded volume interactions. It has been confirmed, in addition, that the key microscopic parameter driving the co-solvent-induced polymer chain collapse is the difference between the energetic parameters of attractive interactions 'polymer-solvent' and 'polymer-co-solvent'.

Despite this mean-field model success in its application to the co-nonsolvency description, the theoretical investigation of the pressure effect on co-nonsolvency has not been addressed till now. As was already mentioned above, high-pressure co-nonsolvency suppression was theoretically analyzed only by the MD simulation. The absence of the theoretical analysis of this fascinating phenomenon within the existing analytic models is related to the fact that they deal with the Helmholtz free energy as the solution thermodynamic potential, though a more adequate thermodynamic potential for such kind of systems is the Gibbs free energy.  Moreover, to verify our theoretical model, it is interesting to make a direct comparison of the theoretical results with the results of MD simulations provided in the literature. In the present paper, we address these issues.

\section{Theoretical background}
Let the flexible isolated polymer chain be immersed in a binary mixture of good solvents at certain co-solvent mole fraction $x$, temperature $T$, and pressure $P$. The contour length of the polymer chain is $Nb$ ($N$ is the polymerization degree and $b$ is the bond length). As in our previous works \cite{Budkov_conons,Budkov1,Budkov2,Budkov3} and in works \cite{Dzubiella2013,Odagiri2015}, for convenience the whole solution volume is divided into two sub-volumes - the first is the volume occupied by the polymer chain (gyration volume) and the rest is the volume of the solution (bulk). The gyration volume is chosen to be spherical $V_g = 4 \pi R_g^3/ 3$, where $R_g$ is the gyration radius. The presence of a polymer chain in the solution changes the environment around it, so the local composition of the mixed solvent near the polymer backbone is different from the composition in the bulk. Thereby, it is reasonable to introduce an additional order parameter -- the local co-solvent mole fraction $x_{1}$.  Thus, the change in the Gibbs free energy of the polymer solution (free energy of polymer solvation) is the appropriate thermodynamic potential, so its minimum with respect to the order parameters ($R_{g}$ and $x_{1}$) determines the thermodynamically stable polymer chain conformation.
Thus, the solvation free energy can be written as follows
\begin{eqnarray}
\Delta G_{s}(R_{g},x_{1})=F_{id}(R_{g},x_{1})+F_{ex}(R_{g},x_{1})+PV_{g}-\mu_{s}N_{s}-\mu_{c} N_{c}, \label{eq:DeltaG}
\end{eqnarray}
where $F_{id}$ is the ideal free energy part, $F_{ex}$ is the excess free energy part, $\mu_s$ and $\mu_c$ are, respectively, the chemical potentials of the solvent and co-solvent; $N_s$ and $N_c$ are the numbers of solvent and co-solvent molecules in the gyration volume, respectively. We neglect the contribution from the surface energy of the gyration volume/bulk interface which is unimportant for the not dense globules that are realized in the co-nonsolvency case. In turn, both ideal and excess free energies can be represented as a sum of independent contributions. Namely, the ideal part consists of entropic terms for the solvent, the co-solvent and the polymer chain. The latter is calculated within the Fixman approximation for a flexible polymer chain \cite{Fixman,Budkov2016,Birshtein,Grosberg,Budkov1,Budkov2,Budkov3}. The inter-molecular interactions are modelled by the Lennard-Jones (LJ) potentials. In order to account for the contributions of attractive and repulsive parts of the LJ potentials to the total free energy, we used the Weeks-Chandler-Anderson (WCA) procedure, introducing the effective diameters of hard spheres in accordance with the Barker-Henderson expression \cite{Hansen_MacDonald} (see also the Supporting information). The latter allows us to minimize the difference between the repulsive contributions of the LJ potential and the hard-core potential of the pure components \cite{Hansen_MacDonald}. Thus, the contribution of repulsive interactions is determined through the Mansoori-Charnahan-Straling-Leland (MCSL) equation of state for the three-component mixture of hard spheres with effective diameters \cite{Mansoori}.
To describe qualitatively co-nonsolvency, one can use the Flory-Huggins (FH) equation of state generalized for three-component mixture (see, for instance, \cite{Freed2015,Opferman2012}) instead the adopted here MCSL equation of state. Nevertheless, in contrast to the MCSL equation of state, the FH equation of state does not take into account the difference in the effective diameters of species, though the latter is important for the quantitative description of co-nonsolvency in real polymer solutions.

It should be noted that in the present theory, the number densities of monomers $\rho_{p}$, solvent $\rho_{s}$, and co-solvent $\rho_{c}$ satisfy the following incompressibility condition $\rho_{p}+\rho_{s}+\rho_{c}=\rho(P,T,x)$, where $\rho(P,T,x)$ is the total number density of the bulk mixture, depending on the temperature $T$, pressure $P$, and composition $x$ through the equation of state $P=P(\rho,T,x)$ (see Appendix A). We would like to stress that the mentioned above incompressibility condition is a good approximation for the liquid state region of the mixture (where the isothermal compressibility $\chi_{T}=\left(\rho^{-1}\partial{\rho}/\partial{P}\right)_{T,x}$ is very small) and allows us to reduce the number of the order parameters to two ones -- the local co-solvent mole fraction $x_{1}=\rho_{c}/(\rho_{c}+\rho_{s})$ and the gyration radius $R_{g}$ \cite{Budkov_conons}. In order to obtain the polymer chain conformational behavior, we minimize the solvation free energy (\ref{eq:DeltaG}) with respect to the gyration radius $R_{g}$ and local composition $x_{1}$.

It is instructive to discuss the connection between our approach and Flory theory for a single polymer chain in a good solvent. As it can be shown (see Supporting information), at $6R_{g}^2/(Nb^2) \gg 1$ and $x_{1}\simeq x$ the Gibbs free energy can be simplified to the standard Flory formula for the polymer chain free energy, so its minimization yields the classical Flory scaling result for the gyration radius $R_{g}\sim N^{3/5}$. The comprehensive explanation of the model and its connection with the Flory theory are given in the Appendix B of this paper.

\section{Numerical results and discussion}
Following the papers \cite{Mukherji2013,Kremer2014,Mukherji2015}, we model the interactions between the molecules of the binary mixture and between the solvent molecules and monomers by the WCA potentials, assuming that $\varepsilon_{p}=\varepsilon_{s}=\varepsilon_{c}=\varepsilon_{ps}=1.0\varepsilon$, $\sigma_{p}=1.0\sigma$, $\sigma_{s}=\sigma_{c}=\sigma_{sc}=0.5\sigma$, and $\sigma_{ps}=0.5\sigma$. We model the interactions between the co-solvent molecules and monomers by the full LJ potential with the interaction parameters $\sigma_{pc}=0.75\sigma$ and $\varepsilon_{pc}=1.0\varepsilon$. For simplicity, we also introduce the reduced temperature $\tilde{T}=k_{B}T/\varepsilon$ and pressure $\tilde{P}=P\sigma^3/\varepsilon$. As in works \cite{Kremer2014,Mukherji2015}, we use the following bond length value $b \approx 0.95\sigma$.

\begin{figure}
\center{\includegraphics[width=0.6\linewidth]{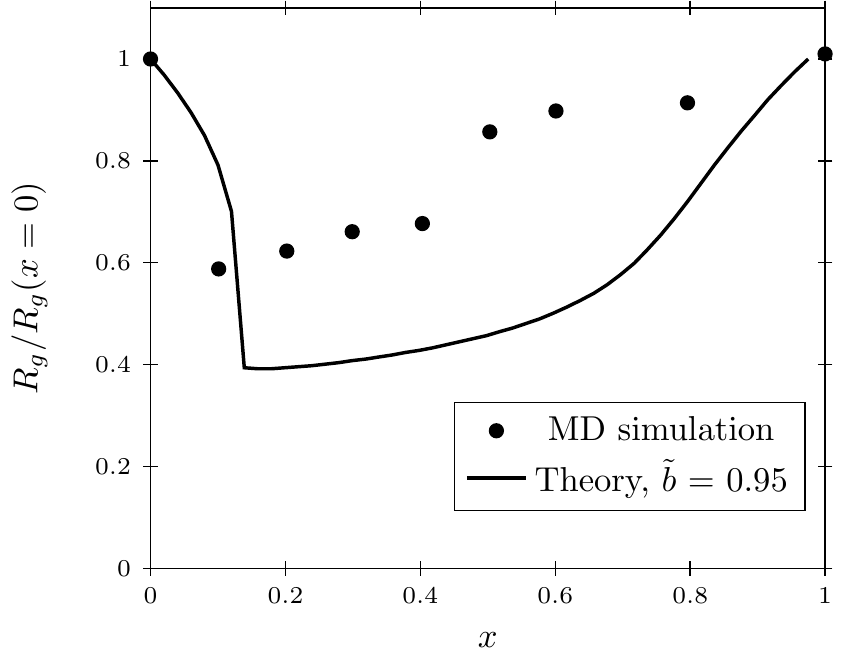}}
\caption{\sl Reduced gyration radius as a function of the co-solvent mole fraction $x$ plotted for $N=30$. The data is shown for $\tilde{T}=0.5$, $\tilde{P}=40$, $\tilde{b}=b/\sigma=0.95$.}
\label{Comp_MD_1}
\end{figure}

\begin{figure}
\center{\includegraphics[width=0.6\linewidth]{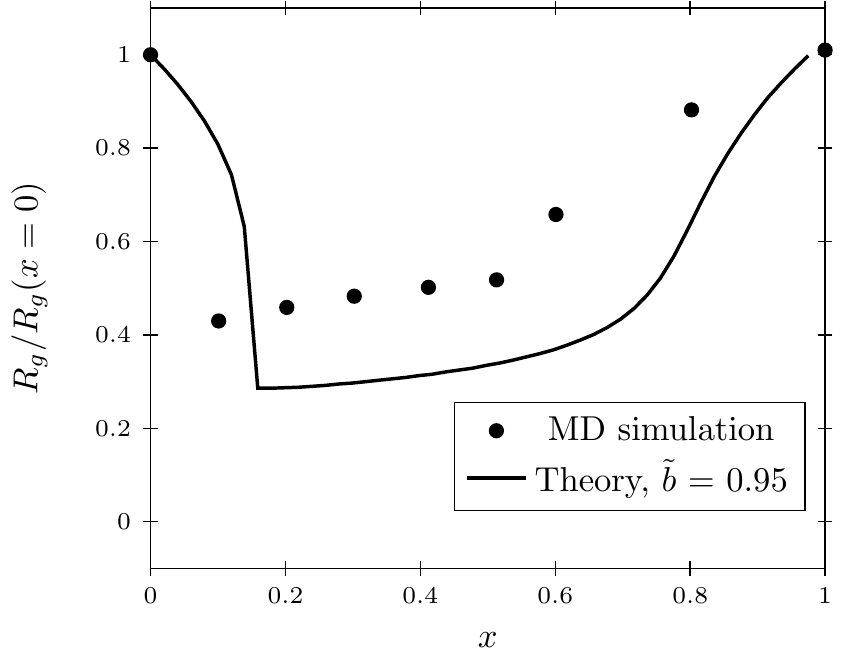}}
\caption{\sl Reduced gyration radius as a function of the co-solvent mole fraction $x$ plotted for $N=100$. The data is shown for $\tilde{T}=0.5$, $\tilde{P}=40$, $\tilde{b}=b/\sigma=0.95$.}
\label{Comp_MD_2}
\end{figure}

At first, we directly compare the prediction of our theoretical model with the MD simulation results presented in works \cite{Kremer2014,Mukherji2015}. Fig. \ref{Comp_MD_1} (for $N=30$) and Fig. \ref{Comp_MD_2} (for $N=100$) demonstrate the reduced gyration radius $R_{g}(x)/R_{g}(x=0)$ as the function of co-solvent mole fraction $x$ at fixed temperature $\tilde{T}=0.5$ and pressure $\tilde{P}=40$ calculated according to the present theory and by the MD computer simulation \cite{Kremer2014,Mukherji2015}. It should be noted that the strength of interaction chosen for the co-solvent selectivity with respect to the monomer is $\varepsilon_{mc}=2k_{B}T$. As is seen, our mean-field model predicts lower gyration radius values in the collapse region than those predicted by MD simulation and shows a good agreement in the regimes of expanded coil conformation. It is worth noting that our mean-field model should describe better conformational behavior of the chains having a rather high degree of polymerization (that guarantees a rather big gyration volume). It explains the fact that the theory shows a better agreement with the MD simulations for the chain with $N=100$ than for $N=30$.

\begin{figure}
\center{\includegraphics[width=0.6\linewidth]{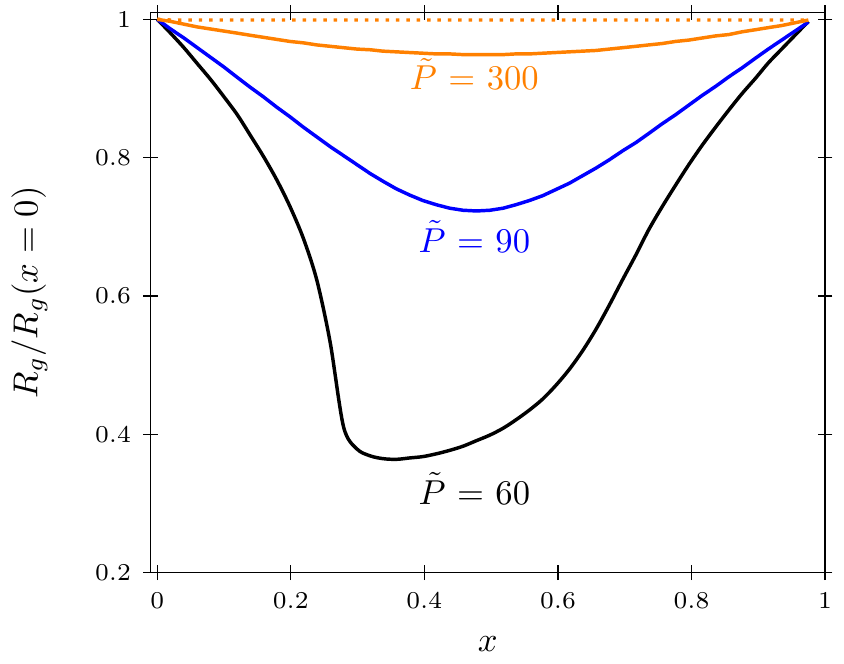}}
\caption{\sl Reduced gyration radius as a function of the co-solvent mole fraction $x$ plotted for different pressure values. The data is shown for $\tilde{T}=0.5$, $\tilde{b}=b/\sigma=0.95$, $N=100$.}
\label{Pressure_eff1}
\end{figure}

\begin{figure}
\center{\includegraphics[width=0.6\linewidth]{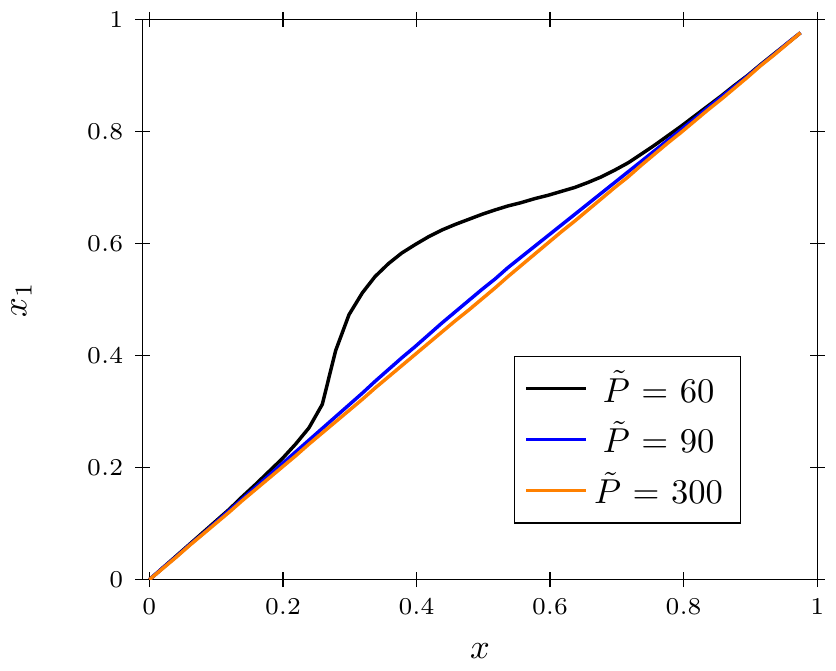}}
\caption{\sl Local co-solvent mole fraction $x_1$ as a function of the co-solvent mole fraction $x$ plotted for different pressure values. The data is shown for $\tilde{T}=0.5$, $\tilde{b}=b/\sigma=0.95$, $N=100$.}
\label{Pressure_eff2}
\end{figure}

In what follows, we investigate the pressure influence on co-nonsolvency. As it was pointed out above, in experiment \cite{Hofmann2} pressure enhancement to the extremely high values destroys co-nonsolvency. As is seen in Fig. \ref{Pressure_eff1}, at rather small pressures, an increase in the co-solvent mole fraction leads to a reentrant coil-globule-coil transition. However, the pressure increase leads to less pronounced minima on the gyration radius curves, so in the region of very high pressures, the coil conformation of the polymer chain becomes almost independent of the co-solvent mole fraction. To understand deeply the nature of this trend, we plot the local co-solvent mole fraction $x_{1}$ in the internal polymer volume depending on the co-solvent mole fraction in the bulk $x$, corresponding to the same pressure values as in Fig. \ref{Pressure_eff1}. Fig. \ref{Pressure_eff2} shows that at rather low pressure the chain collapse is accompanied by the increase in the co-solvent mole fraction $x_1$ in the internal polymer volume (see also \cite{Budkov_conons}). However, the rather high pressure imposed suppresses the co-solvent concentration enhancement in the internal polymer volume and, simultaneously, the polymer chain collapse. Such a behavior of the local co-solvent concentration may indicate on the suppression of the preferential solvation of the polymer backbone by co-solvent, when the pressure increases. The latter is in agreement with the speculations based on the full atomistic MD simulations of PNIPAM in aqueous methanol presented in work \cite{Kremer2015}. Strictly speaking, to relate the high pressure co-nonsolvency suppression to suppression of the preferential solvation, it is necessary to perform an analysis of the radial distribution function 'monomer-co-solvent' at different pressures imposed. However, such analysis beyond the mean-field theory and might be provided by the  computer simulations (MD or Monte-Carlo) or classical density functional theory (DFT). Nevertheless, one can understand what is the main reason of co-nonsolvency suppression at high pressures. As is seen in Fig. \ref{Pressure_eff1}, the gyration radius at high pressures is very close to the values for the chain not attracting the molecules of mixture (in that case $R_{g}/R_{g}(x=0)=1$ at any pressures). It means that the mixture becomes so dense, that the monomers do not anymore feel the attractive interaction with the solvent. This quite trivial interpretation is in agreement with classic result of the liquid state theory that a thermodynamic behavior of the dense liquids must be determined predominantly by the excluded volume of molecules \cite{Hansen_MacDonald}.

A possibility to describe the co-nonsolvency suppression by very high pressure within the present model, taking into account only the universal Van der Waals and excluded volume interactions, indicates that the latter is a generic effect, as co-nonsolvency itself \cite{Mukherji2015}. In other words, the reduction of the LCST by pressure enhancement can be described at the generic level without introducing the hydrogen bonding between monomers and solvent molecules (see ref. \cite{Bekiranov_1993}). Thus, one can expect the co-nonsolvency suppression by high pressure in such mixtures as polystyrene/cyclohexane/dimethylformamide \cite{Wolf78}, where the association between species due to the hydrogen bonding is fully absent.

It is instructive to estimate in physical units the pressure at which co-nonsolvency fades away. Assuming the temperature $T=300~K$, the linear size $\sigma=0.3~nm$, and the dimensionless pressure $\tilde{P}=300$, we obtain the pressure value $P\sim 10^3~MPa$. This value of order the experimental pressure values (see refs. \cite{Hofmann1,Osaka2012}) at which co-nonsolvency disappears in mixture PNIPAM/water/methanol.

\begin{figure}
\center{\includegraphics[width=0.6\linewidth]{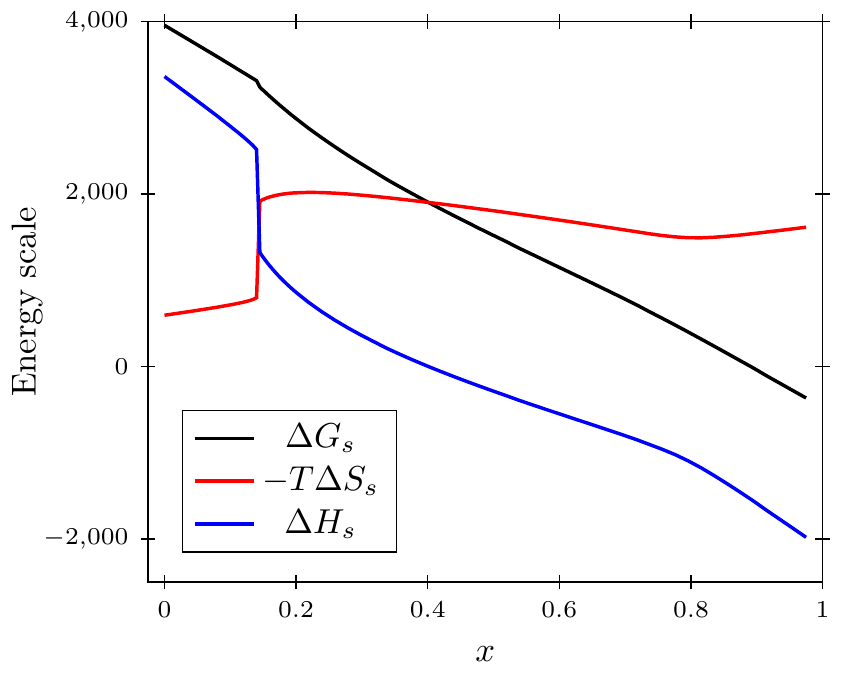}}
\caption{\sl Solvation free energy and its enthalpy and entropy contributions as the functions of the co-solvent mole fraction in the co-nonsolvency region. The data is shown for $N=100$, $\tilde{T}=0.5$, $\tilde{P}=40$.}
\label{Entr_enth}
\end{figure}

Now we turn to discussion of the thermodynamic functions behavior in the co-nonsolvency region. Fig. \ref{Entr_enth} demonstrates the solvation free energy $\Delta G_{s}$ of the polymer chain, as well as its entropy $\Delta S_{s}=\partial{\Delta G_{s}}/\partial{T}$ and enthalpy $\Delta H_{s}=\Delta G_{s}+T \Delta S_{s}$ contributions as the functions of co-solvent mole fraction $x$. As is seen, the solvation free energy decreases continuously with the increase in the co-solvent mole fraction and has an inflection point at the chain collapse point.  Nevertheless, the entropy $T\Delta S_{s}$ and enthalpy $\Delta H_{s}$ contributions abruptly decrease at this point. Such behavior of the thermodynamic functions can be easily interpreted. Indeed, due to the above mentioned preferential solvation, the contribution of attractive interaction 'polymer-co-solvent' to the enthalpy grows in its absolute value, when the chain collapse occurs. The latter leads to an enthalpy decrease. On the other hand, the chain collapse results in a configurational entropy decrease due to the decrease in the free volume available for the monomers and solvent/co-solvent molecules. At a further increase in the co-solvent mole fraction, the enthalpy decreases monotonically, whereas the entropy, on the contrary, remains almost constant. Thus, the theory indicates on the leading role of the enthalpy in the concentration region, where the co-solvent preferential binding with the polymer backbone takes place.

\section{Concluding remarks}
In this short communication, we have demonstrated the applicability of our self-consistent field theory to describing the pressure effect on co-nonsolvency. We have demonstrated that our theory can successfully describe the co-nonsolvency suppression by the pressure enhancement. Analysing the local co-solvent concentration behavior at different pressures, we have speculated that the latter phenomenon is related to the suppression of preferential solvation of the polymer backbone by high pressure imposed. We have obtained a good agreement between our theoretical results and MD simulation results \cite{Kremer2014,Mukherji2015} for the co-solvent-induced reentrant coil-globule-coil transition of the Lennard-Jones flexible chain in the binary mixture of the Lennard-Jones fluids. We have shown that the co-solvent-induced polymer chain collapse is accompanied by an abrupt decrease in the entropy and enthalpy contributions to the solvation free energy, although the latter remains continuous.

Due to the fact that in present research we aimed to compare directly the model predictions with the MD simulation results of Mukherji et al. \cite{Mukherji2015}, rather than with the experimental data, we have neglected the attractive interactions 'solvent-solvent', 'solvent-co-solvent', 'solvent-monomer', 'co-solvent-co-solvent', and 'monomer-monomer'. However, to apply our model to treating the available experimental data (for instance, for PNIPAM in aqueous methanol), it is necessary, in general case, to take into account all types of the attractive interactions \cite{Budkov_conons}. Moreover, in this model an association between species caused by hydrogen bonding is also neglected. The latter might be accounted for in the same manners as in refs. \cite{Bekiranov_1993,Freed2016,Tamm2002,Tanaka2008} We would also like to stress that the present model, being in its nature an off-lattice model, is based on the modern liquid-state theory \cite{Hansen_MacDonald}. Thereby, the microscopic interaction parameters, taking place in this model, cannot be related directly to those are in the FH-type lattice models (FH parameters and excluded volume parameter). Nevertheless, it is interesting to compare an applicability of the present off-lattice liquid-state model \cite{Budkov_conons}, three-component Flory-Huggins model \cite{Shild,Freed2015}, and adsorption model formulated by Mukherji et al. \cite{Mukherji2015,Kremer2014} to available experimental data that could be a subject of the future publications.

\begin{acknowledgments}

\end{acknowledgments}
The authors thank Oleg Borisov for drawing their attention to some inaccuracies in the numerical calculations. The authors thank D. Mukherji for valuable comments and discussions. The authors thank anonymous Referees for valuable comments and remarks allowed us improve the text. The research was prepared within the framework of the Academic Fund Program at the National Research University Higher School of Economics (HSE) in 2017-2018 (grant No 17-01-0040) and by the Russian Academic Excellence Project "5-100". The authors contributed equally to the present research.

\section{Appendix}
\subsection{Appendix A: Solvation free energy and its minimization procedure}
Here we present the details of theoretical model omitted in the main text.
We start from the solvation free energy of the polymer chain in the mixed solvent media
\begin{equation}
\nonumber
\Delta G_{s}=F_{id}+F_{ex}+PV_{g}-\mu_{s}N_{s}-\mu_{c} N_{c},
\end{equation}
where $V_{g}=4\pi R_{g}^3/3$ is the volume of gyration of the polymer chain, $R_{g}$ is the chain gyration radius,  $N_{s}$ and $N_{c}$ are, respectively, the molecule numbers of the solvent and co-solvent in the gyration volume; $F_{id}$ is the ideal free energy of the polymer chain and mixed solvent which can be calculated in the following way
\begin{equation}
\label{eq:Fid}
F_{id}=\frac{9}{4}k_{B}T\left(\frac{6R_{g}^2}{Nb^2}+\frac{Nb^2}{6R_{g}^2}\right)\nonumber
\end{equation}
\begin{equation}
+N_{s}k_{B}T\left(\ln{\frac{N_{s}\Lambda_{s}^3}{V_{g}}}-1\right)+N_{c}k_{B}T\left(\ln{\frac{N_{c}\Lambda_{c}^3}{V_{g}}}-1\right),
\end{equation}
where $b$ is the bond length of the chain, $k_{B}$ is the Boltzmann constant, $N$ is the polymerization degree, $T$ is the absolute temperature, $\Lambda_{s}$ and $\Lambda_{c}$ are the de Broglie wavelengths of the low-molecular weight species. The first term in (\ref{eq:Fid}) is the free energy of the ideal Gaussian polymer chain within the Fixman approximation \cite{Fixman,Budkov2016,Birshtein,Grosberg,Budkov1,Budkov2,Budkov3,Budkov4,Budkov5,Brilliantov1998,Tom2016,Tom2017}; $P$ is the pressure imposed to the system which will be determined below.
The interactions 'monomer-monomer', 'monomer-solvent', 'solvent-solvent', 'co-solvent-co-solvent',  and 'solvent-co-solvent' are described by the WCA potentials
\[
    V_{ij}(r)=\left\{
                \begin{array}{ll}
                  4\epsilon_{ij}\left[\frac{1}{4}+ \left(\frac{\sigma_{ij}}{r}\right)^{12} -  \left(\frac{\sigma_{ij}}{r}\right)^{6}\right], r<2^{1/6}\sigma_{ij}\\
                  0, r>2^{1/6}\sigma_{ij}\\
                \end{array}
              \right\}.
  \]

Interaction monomer-co-solvent is described by the full Lennard-Jones potential
\begin{equation}
V_{pc}(r)= 4\epsilon_{pc}\left[ \left(\frac{\sigma_{pc}}{r}\right)^{12} -  \left(\frac{\sigma_{pc}}{r}\right)^{6}\right].
\end{equation}

Therefore, the excess free energy of polymer solution takes the form
\begin{equation}
\label{eq:ev}
F_{ex}=F_{ev}+F_{att},
\end{equation}
where $F_{ev}$ is the contribution of the repulsive interactions in the gyration volume due to the excluded volume of the monomers and molecules of the low-molecular weight species which we determine through the Mansoori-Carnahan-Starling-Leland equation of state for the hard-spheres mixture (see below) with the effective diameters of species calculated in accordance with a well-known Barker-Henderson relation \cite{Hansen_MacDonald}:
\begin{equation}
\label{diameters}
d_{i} = \int_0^{2^{1/6}\sigma_{ii}} \left(1 - e^{-V_{ii}(r)/k_B T}\right) dr,
\end{equation}
where $i=p,s,c$.

As it was mentioned in the main text, our model is fully corresponded to situation realized in MD simulation of Mukherji et al. \cite{Mukherji2013}. Thereby, we neglected the attractive interactions 'solvent-solvent', 'solvent-co-solvent', 'solvent-monomer', 'co-solvent-co-solvent', and 'monomer-monomer', taking into account attractive interaction only between polymer and co-solvent within the standard mean-field approximation:
\begin{equation}
F_{att} =  \rho_p \rho_c V_{g} \int d\bold{r}   \Phi_{pc} (|\bold{r}|)=-\frac{32}{9}\sqrt{2}\pi\epsilon_{pc}\sigma^3_{pc}\rho_p\rho_c V_{g},
\end{equation}
where $V_g=4\pi R_{g}^3/3$ is the gyration volume, $\rho_{p}$ and $\rho_{c}$ are, respectively, the number densities of monomers and co-solvent in the gyration volume; attractive part of the full Lennard-Jones potential according to the Weeks-Chandler-Anderson scheme \cite{Hansen_MacDonald} is
\[
    \Phi_{pc}(r)=\left\{
                \begin{array}{ll}
                  -\epsilon_{pc}, r<2^{1/6}\sigma_{pc}\\
                  4\epsilon_{pc}\left[ \left(\frac{\sigma_{pc}}{r}\right)^{12} -  \left(\frac{\sigma_{pc}}{r}\right)^{6}\right], r>2^{1/6}\sigma_{pc}\\
                \end{array}
              \right.
  \].

Choosing the local mole fraction of co-solvent $x_{1}$ in the gyration volume and the gyration radius $R_{g}$ as the order parameters, one can rewrite the solvation free energy in the following way
\begin{equation}
\label{eq:DeltaG2}
\Delta G_{s}(R_{g},x_{1})=\frac{9}{4}k_{B}T\left(\frac{6R_{g}^2}{Nb^2}+\frac{Nb^2}{6R_{g}^2}\right)\nonumber
\end{equation}
\begin{equation}
+\rho_{1}V_{g}k_{B}T\left(x_{1}\left(\ln\left(\rho_{1}x_{1}\Lambda_{c}^3\right)-1\right)+(1-x_{1})\left(\ln\left(\rho_{1}(1-x_{1})\Lambda_{s}^3\right)-1\right)\right)\nonumber
\end{equation}
\begin{equation}
+V_{g}\left(P(\rho,x,T)+f_{ex}(\rho,x_{1},\rho_{p},T)-\rho_{1}\left(\mu_{s}(\rho,x,T)(1-x_{1})+\mu_{c}(\rho,x,T)x_{1}\right)\right),
\end{equation}
where $\rho_{p}=N/V_{g}$ is a monomer number density and $f_{ex}(\rho,x_{1},\rho_{p},T)$ is a density of excess free energy which has a form
\begin{equation}
f_{ex}(\rho,x_{1},\rho_{p},T)=\rho k_{B}T A(\rho,x_{1},\rho_{p})-\frac{32}{9}\sqrt{2}\pi\epsilon_{pc}\sigma^3_{pc}\rho_p\rho_1 x_1,
\end{equation}
where the following short-hand notations are introduced
\begin{equation}
A(\rho,x_{1},\rho_{p})=-\frac{3}{2}\left(1-y_1(\rho,x_{1},\rho_{p})+y_2(\rho,x_{1},\rho_{p})+y_3(\rho,x_{1},\rho_{p})\right)+\frac{3y_2(\rho,x_{1},\rho_{p})+2y_3(\rho,x_{1},\rho_{p})}{1-\xi(\rho,x_{1},\rho_{p})}\nonumber
\end{equation}
\begin{equation}
+\frac{3\left(1-y_1(\rho,x_{1},\rho_{p})-y_2(\rho,x_{1},\rho_{p})-\frac{y_3(\rho,x_{1},\rho_{p})}{3}\right)}{2(1-\xi(\rho,x_{1},\rho_{p}))^2}+(y_3(\rho,x_{1},\rho_{p})-1)\ln(1-\xi(\rho,x_{1},\rho_{p})),
\end{equation}
\begin{equation}
y_{1}(\rho,x_{1},\rho_{p})=\Delta_{cp}\frac{d_{c}+d_{p}}{\sqrt{d_{p}d_c}}+\Delta_{sp}\frac{d_{s}+d_{p}}{\sqrt{d_{p}d_s}}+\Delta_{sc}\frac{d_{s}+d_{c}}{\sqrt{d_{c}d_s}},
\end{equation}
\begin{equation}
y_2(\rho,x_{1},\rho_{p})=\frac{1}{\xi}\left(\frac{\xi_c}{d_c}+\frac{\xi_s}{d_s}+\frac{\xi_{p}}{d_{p}}\right)\left(\Delta_{cp}\sqrt{d_c d_p}+\Delta_{sp}\sqrt{d_s d_p}+\Delta_{sc}\sqrt{d_{s} d_{c}}\right),
\end{equation}
\begin{equation}
y_{3}(\rho,x_{1},\rho_{p})=\left(\left(\frac{\xi_{c}}{\xi}\right)^{2/3}\left(\frac{\rho_{1}x_{1}}{\rho}\right)^{1/3}+\left(\frac{\xi_{s}}{\xi}\right)^{2/3}\left(\frac{\rho_{1}(1-x_{1})}{\rho}\right)^{1/3}+\left(\frac{\xi_{p}}{\xi}\right)^{2/3}\left(\frac{\rho_{p}}{\rho}\right)^{1/3}\right)^{3},
\end{equation}
\begin{equation}
\label{eq:Delta}
\Delta_{sp}=\frac{\sqrt{\xi_{s}\xi_{p}}}{\xi}\frac{(d_{s}-d_{p})^2}{d_{s}d_{p}}\frac{\sqrt{\rho_{1}\rho_{p}(1-x_{1})}}{\rho},~\Delta_{cp}=\frac{\sqrt{\xi_{c}\xi_{p}}}{\xi}\frac{(d_{c}-d_{p})^2}{d_{c} d_{p}}\frac{\sqrt{\rho_{1}\rho_{p}x_{1}}}{\rho},
\end{equation}
\begin{equation}
\label{eq:Deltacs}
\Delta_{cs}=\frac{\sqrt{\xi_{c}\xi_{s}}}{\xi}\frac{(d_{c}-d_{s})^2}{d_{c}d_{s}}\frac{\rho_{1}}{\rho}\sqrt{x_{1}(1-x_{1})}
\end{equation}
\begin{equation}
\xi_{s}=\frac{\pi \rho _{1}(1-x_{1})d_{s}^3}{6}, ~ \xi_{c}=\frac{\pi  \rho _{1}x_{1}d_{c}^3}{6}, ~ \xi_{p}=\frac{\pi \rho_{p}d_{p}^3}{6},
\end{equation}
\begin{equation}
\xi=\xi(\rho,x_{1},\rho_{p})=\xi_{s}+\xi_{c}+ \xi_{p};
\end{equation}
the local solvent composition $x_{1}$ in the gyration volume is introduced by the following relations
\begin{equation}
\rho_{s}=\frac{N_{s}}{V_{g}}=\rho_{1}(1-x_{1}),~\rho_{c}=\frac{N_{c}}{V_{g}}=\rho_{1}x_{1}.
\end{equation}
The local number density $\rho_{1}$ of binary mixture can be related with the bulk number density $\rho$ and the monomer number density $\rho_{p}$ through the incompressibility condition $\rho_{1}=\rho - \rho_{p}$.

The pressure in the bulk solution $P$ in our model is determined by the the Mansoori-Carnahan-Starling-Leland equation of state:
\begin{equation}
\label{eq:P}
\frac{P(\rho,x,T)}{\rho k_{B}T}= \frac{1+\xi(\rho,x,0)+\xi^2(\rho,x,0)-3\xi(\rho,x,0)(y_1(\rho,x,0)+y_2(\rho,x,0)\xi(\rho,x,0)+\frac{\xi^2(\rho,x,0)y_3(\rho,x,0)}{3})}{(1-\xi(\rho,x,0))^3}.
\end{equation}
The chemical potentials of the solvent species can be calculated by the following obvious thermodynamic relations
\begin{equation}
\label{eq:muc}
\mu_{c}(\rho,x,T)=\frac{1}{\rho}\left(P(\rho,x,T)+f(\rho,x,T)+(1-x)\left(\frac{\partial{f(\rho,x,T)}}{\partial{x}}\right)_{\rho,T}\right),
\end{equation}
\begin{equation}
\label{eq:mus}
\mu_{s}(\rho,x,T)=\frac{1}{\rho}\left(P(\rho,x,T)+f(\rho,x,T)-x\left(\frac{\partial{f(\rho,x,T)}}{\partial{x}}\right)_{\rho,T}\right),
\end{equation}
where $f(\rho,x,T)$ is a density of Helmholtz free energy of the bulk solution which can be calculated as
\begin{equation}
\label{eq:f}
f(\rho,x,T)=\rho k_{B}T\left(x\ln\left(\rho\Lambda_{c}^3 x\right)+(1-x)\ln\left(\rho\Lambda_{s}^3 (1-x)\right)\right)+\rho k_{B}T A(\rho,x,0).
\end{equation}

\subsection{Appendix B: Connection with Flory theory}
Here we present how our approach can be related to the classic Flory theory of a single flexible polymer chain in a good solvent. We rewrite the Gibbs free energy as follows:
\begin{equation}
\Delta G_{s}=\frac{9}{4}k_{B}T\left(\frac{6R_{g}^2}{Nb^2}+\frac{Nb^2}{6R_{g}^2}\right)+V_{g}\left(f_{mix}+P-\mu_{s}\rho_{1}(1-x_{1})-\mu_{c}\rho_{1}x_{1}\right),
\end{equation}
where $f_{mix}$ is the free energy density of three-component mixture of unbound particles.
We consider the regime of expanded coil, i.e., when $6R_{g}^2/(Nb^2)\gg 1$ and $x_{1}\simeq x$. In this case, the internal monomer number density is small, i.e $\rho_{p}\ll \rho$, so that $\rho_{1}\simeq \rho$. Hence, we get in this approximation
\begin{equation}
f_{mix}(\rho,x_{1},\rho_{p},T)=f(\rho,x,T)+\frac{1}{2}B(\rho,x,T)\rho_{p}^2 +O(\rho_{p}^3),
\end{equation}
where the second virial coefficient
\begin{equation}
B(\rho,x,T)=\frac{\partial^2 f_{mix}(\rho,x,0,T)}{\partial\rho_{p}^2}
\end{equation}
is introduced and $f(\rho,x,T)$ is determined by (\ref{eq:f}).
Further, taking into account that $f+P-\mu_{s}\rho(1-x)-\mu_{c}\rho x=0$, we arrive at the relation for the single chain free energy \cite{Khohlov} which depends on the state parameters of solvent mixture only through the second virial coefficient of monomers $B$:
\begin{equation}
\Delta G_{s}=F_{p}(R_{g})=\frac{9}{4}k_{B}T\left(\frac{6R_{g}^2}{Nb^2}+\frac{Nb^2}{6R_{g}^2}\right)+\frac{B(\rho,x,T)N^2}{2V_{g}}.
\end{equation}
Minimization of the polymer free energy with respect to the gyration radius yields the classic Flory result
\begin{equation}
R_{g}\sim b^{2/5}B^{1/5}N^{3/5}.
\end{equation}

\newpage

\end{document}